\documentclass{pinchcr}   

\usepackage{amssymb}
\usepackage{amsmath}

\usepackage[dvips]{graphicx}

\title{Defects on Cylinders:  Superfluid Helium Films and Bacterial Cell Walls}

\author{David R. Nelson and Ariel Amir}
\affiliation{Lyman Laboratory of Physics, Harvard University, Cambridge, MA 02138, USA}
\authors{2}

\begin{document}

\maketitle

%

\preface

\noindent There is a deep analogy between the physics of crystalline solids and the behavior of superfluids, dating back to pioneering work of Phillip Anderson, Paul Martin and others.   The stiffness to shear deformations in a periodic crystal resembles the superfluid density that controls the behavior of supercurrents in neutral superfluids such as He${}^{4}$.  Dislocations in solids have a close analogy with quantized vortices in superfluids.    Remarkable recent experiments on the way rod-shaped bacteria elongate their cell walls have focused attention on the dynamics and interactions of point-like dislocation defects in partially ordered cylindrical crystalline monolayers.   In these lectures, we review the physics of superfluid helium films on cylinders and discuss how confinement in one direction affects vortex interactions with supercurrents.   Although there are similarities with the way dislocations respond to strains on cylinders, important differences emerge, due to the vector nature of the topological charges characterizing the dislocations.

\tableofcontents

\maintext

\chapter {Introduction}

 Defects of various types have long been known to play a key role in condensed matter physics and materials science.    Point-like vacancy and interstitial defects in crystalline solids facilitate atomic diffusion and give rise to color centers \cite{bib1}.   Vortex lines in type II superconductors, singularities that concentrate magnetic field lines, move in the presence of an external electrical current  \cite{bib2,bib3}, causing dissipation unless the vortices are strongly pinned, an especially difficult problem for practical applications of  high temperature superconductors, due to thermal fluctuations  \cite{bib4}.   Soft metals bend irreversibly when a dilute concentration of translational line defects called dislocations (see below) move in response to an external stress  \cite{bib5}.    The Bronze Age began when early metallurgists added small concentrations of tin to copper to create a stronger and more effective material --  the tin impurities pin dislocations in place, thus preventing plastic deformation.  The Iron Age was the result of making iron even stronger than bronze by supplementing the addition of impurities (typically carbon) with work hardening at the blacksmith's forge, which entangles the dislocation lines  \cite{bib6}.   Finally, melting of two-dimensional crystalline solids with increasing temperature can be driven by the serial unbinding of thermally activated point-like dislocations and then point-like rotational defects called disclinations, with an intervening hexatic phase  \cite{bib7}.   The detailed theory found early confirmation with experiments on well-equilibrated 2d assemblies of colloidal particles  \cite{bib8}, and most recently in a remarkable decade-long sequence of experiments on super-paramagnetic particles interacting with a repulsive $1/r^{3} $ potential by the group of Georg Maret at the University of Konstanz  \cite{bib9}.

\begin{figure}
\includegraphics[width=0.9 \textwidth]{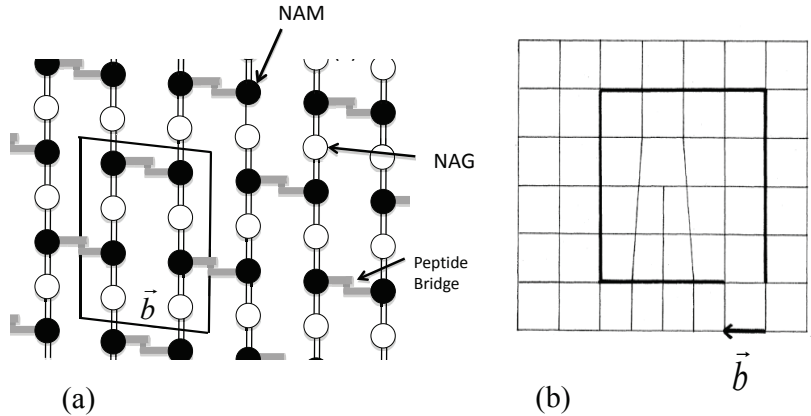}
\vspace{-0.5cm}
\caption{ (a) Peptidoglycan layer of a bacterial cell wall. Glycan strands of alternating NAM and NAG sugars are connected by short peptide linkages. (b) An edge dislocation arising from the insertion of a extra half-row of unit cells like that in (a) ; as the dislocation moves upwards, it mediates easy extension of a glycan free end and is characterized by the Burgers vector $\vec{b}$, the amount by which a circuit that would close in a perfect crystal fails to close due to the defect.
}
\label{fig1}
\end{figure}

Remarkably, dislocation dynamics \textit{also} plays a significant role in the two-dimensional covalently bonded macromolecules that comprise the cell walls of $\sim$ 1 micron-diameter rod-shaped bacteria. A fragment of such a cell wall is shown in Fig. \ref{fig1}a, consisting of a regular meshwork of glycan (\emph{i.e.}, sugar) strands of alternating moieties of NAM (N-acetylmuranimic acid) and NAG (N-acetylglucosamine), with these chains connected by short peptide (\emph{i.e.}, amino acid) crosslinks  \cite{bib10}. This peptidoglycan meshwork ranges from a single layer in some thin-walled gram-negative bacteria to of order 8-15 layers in the thicker cell walls of gram positive organisms. Because of its nonzero shear modulus, the cell wall maintains the cylindrical shape of the bacteria, as well as resisting anywhere from 1-20 atmospheres of outward osmotic pressure. If the cell wall is gently eliminated (and rupture is avoided), rod-shaped bacteria become spherical, due to the fluid character of the confinement by the remaining lipid bilayers. While carrying out their important functions, cell walls must maintain their integrity as bacteria such as \emph{E. coli} elongate to twice their initial size approximately every 20-30 minutes, followed by cell division. It is as if a pressurized dirigible aircraft had to be systematically remodeled to gradually become twice its initial size while still in flight!

\begin{figure}
\includegraphics[width=0.4\textwidth, angle=-90]{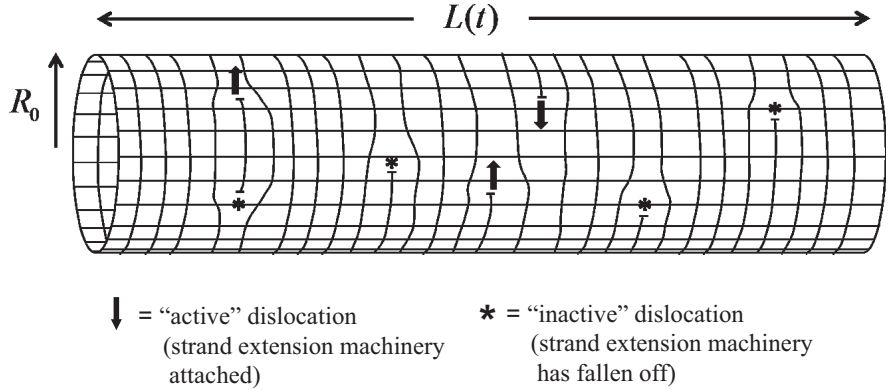}

\caption{ Schematic of dislocations  on the cylindrical portion of a bacterial cell wall. Circumferentially rotating “activated” dislocations propelled by attached strand elongation machinery are indicated by arrows. “Inactive” dislocations are indicated by asterisks. In reality, the inactivated dislocations are likely to greatly outnumber the activated ones. Here, $R_0$ is the constant radius the cylindrical cell wall of the bacteria, and $L(t)$ is its length, which grows as dislocation motion causes the bacterium to elongate.
}
\label{fig2}
\end{figure}

It turns out that point-like edge dislocation defects provide a natural way for two-dimensional partially ordered bacterial cell walls to remodel themselves.  As illustrated schematically in Fig. \ref{fig1}b, an edge dislocation can be regarded as an extra half-row of unit cells  \cite{bib5,bib11}.   Dislocations are characterized by a topological charge called the Burgers vector $\vec{b}$, given by the amount the contour integral of the displacement field $\vec{u}(x,y)$around the defects fails to close,
\begin{equation} \label{GrindEQ__1_}
\oint d\vec{u} =\vec{b}.
\end{equation}
If the extra row of unit cells terminating in dislocations like those in Fig. \ref{fig2} could be systematically extended upwards by adding new material, the crystal would widen along the axis of the cylinder.  The cylinder will elongate at constant radius as a subset of ``activated'' dislocations circumnavigates via this climbing motion and extends glyan strands.   This defect mechanism for bacterial elongation, with the activated defects called  ``glycan extension centers'', was proposed by Burman and Park in 1984  \cite{bib12,bib13}, approximately 50 years after the importance of dislocations was recognized in materials science  \cite{bib14,bib15,bib16}.  From a materials science perspective, it is at first hard to see how dislocation dynamics could ever be relevant in networks of the strong covalent bonds that characterize a bacterial cell wall -- dislocations were invented to understand plastic deformations of metals, where atoms are connected by much weaker metallic bonds.   How can the very strong bonds in Fig. \ref{fig1}a break and reform to remodel the cell wall at physiological temperatures?   Biology accomplishes this seemingly impossible task because it has evolved a specialized suite of enzymatic molecular machines with names such as MreB, RodA, PBP1b, PBP2, MreC and MreD  \cite{bib17}.     There is a rough analogy between this collection of enzymes and the enzymes that allow DNA replication every time a cell divides  \cite{bib12}.    Instead of facilitating the complementary base pairing necessary to complete a single strand of DNA, these enzymes remodel cell walls by breaking peptide crosslinks, extending glycan strands and then adding new amino acid cross-bridges.   Interestingly, only $\sim$ 20-30 out of the approximately 10,000 glycan strand ends in a typical bacterial cell wall  \cite{bib13} need to be activated to allow a cell to double its length in about 30 minutes  \cite{bib12}.   Pairs of broken strand ends are associated with the end of a row of unit cells like those in Fig. \ref{fig1}a, and thus qualify as a dislocation, according to the criterion illustrated in Fig. \ref{fig1}b.  The suite of remarkable enzymes that creates an activated dislocation can only lengthen of the cell wall -- new enzymes are needed to form the septum that partitions a growing cell into two daughters and completes a cell division  \cite{bib17}.

In these lecture notes, we will \textit{not} describe in detail the application of dislocation theory to this fascinating biophysics problem.   A detailed exposition can be found in Refs.  \cite{bib18,bib19,bib20}.   Instead, we focus on the simpler, but mathematically closely-related problem of point-like superfluid vortices disrupting order in the condensate wave function of superfluid helium films on cylinders.    Superfluid helium has a rich and illustrious history, enriched by ideas from such giants of theoretical physics as Fritz London, Lars Onsager and Richard Feynman  \cite{bib21,bib22,bib23}.  The deep analogy between the physics of crystalline solids and the behavior of superfluids and superconductors, first fruitfully exploited by scientists such as P. W. Anderson and P. C. Martin  \cite{bib24,bib25}, is summarized for two dimensional systems in Table I.   Here, $\vec{u}(\vec{r})$ is the same phonon displacement field measuring deviations from a perfect crystalline lattice that appears in Eq. \eqref{GrindEQ__1_}, and $\theta (\vec{r})$ describes the phase of the superfluid order parameter.    Its relation of its gradient to the superfluid velocity in terms of Planck's constant $\hbar $ and the mass $m$ of a helium atom will be discussed further below.   The energy of both systems is quadratic in the derivatives of these fields, with coefficients or ``stiffnesses'' given by elastic constants and the superfluid density respectively.   The last row of the table is the quantization condition on point-like defects, given by a Burgers vector $\vec{b}$ equal to a lattice displacement for dislocations in crystalline solids, and by an integer $s$ for vortices in superfluid helium films.

\vspace{0.5cm}

\noindent Table I:   Analogy between the elastic deformations of two-dimensional crystalline solids and low energy excitations of superfluid He${}^{4}$ films.

\vspace{0.2cm}

\noindent
\begin{tabular}{|p{1.4in}|p{1.8in}|p{2.0in}|} \hline
 & Crystalline Solid & Superfluid Helium \\ \hline
 Dynamical field & $\begin{array}{l} {{\rm strain\; matrix}} \\ {u_{ij} (\vec{r})=\frac{1}{2} \left[\partial _{i} u_{j} (\vec{r})+\partial _{j} u_{i} (\vec{r})\right]} \end{array}$  & $\begin{array}{l}{{\rm superfluid\; velocity}} \\ {\vec{v}_{s} (\vec{r})=\frac{\hbar }{m} \vec{\nabla }\theta (\vec{r})} \end{array}$ \\ \hline
Energy & $E=\frac{1}{2} \int d^{2} r\left[2\mu u_{ij}^{2} +\lambda u_{kk}^{2} \right] $  & $U=\frac{1}{2} \rho _{s} \int d^{2} r\left[\frac{\hbar }{m} \vec{\nabla }\theta (\vec{r})\right] ^{2} $  \\ \hline
 Stiffness &  $\mu {\rm \; and\; }\lambda {\rm \; (Lame'\; coefficients)}$  & $\rho _{s} \frac{\hbar ^{2} }{m^{2} } {\rm \; (}\rho _{s} {\rm \; =\; superfluid\; density})$  \\ \hline
Defect contour integral  & $\oint d\vec{u}(\vec{r}) =\vec{b}$  & $\oint d\theta (\vec{r}) =2\pi s{\rm ,\; }s=0,\pm 1,\pm 2,...$ $ $  \\ \hline
\end{tabular}

\noindent

\vspace{0.5cm}

In what follows, we describe the phenomena associated with the right-hand column of Table I, eventually showing how the excitation of quantized vortex pairs triggers the decay of super currents on cylinders, thus extending the earlier work of Ambegaokar et al. on planar helium films  \cite{bib26}. In Sec. II, we review the physics of inviscid superfluid helium films. As we shall see, low energy excitations of these very thin films, which can be less than a 100 atomic layers thick, resemble gravity waves in a classical fluid of finite depth, except that the dominant restoring force is the van der Waals interaction with the substrate.   These van der Waals waves, also called ``third sound'', were predicted and observed experimentally by K. R. Atkins and coworkers approximately 50 years ago  \cite{bib27,bib28}.  Third sound fails to propagate above temperatures exceeding a few degrees above absolute zero  \cite{bib29,bib30}.    In the limit of long-wavelength third sound, corresponding to approximately uniform supercurrents, this failure can be attributed to the excitation of quantized vortex pairs with equal and opposite vorticity. In Secs. III, and IV we show how the energetics of superfluid vortices can be mapped onto two-dimensional electrostatics embedded in a cylinder, with a crossover from a 2d logarithmic potential at short distances to a linear confinement when distances exceed the cylinder circumference.   After discussing the nucleation of vortex pairs on a cylinder in the presence of a mixture of longitudinal and azimuthal supercurrents, in Sec. V we briefly compare and contrast the energetics of the \textit{vector} dislocation charges for two-dimensional crystals wrapped around a cylinder. Charge-neutral dislocation pairs with Burger's vectors at a generic angle to the cylinder axis behave much like their counterparts in scalar 2d electrostatics. However, for the special case of Burgers vectors aligned with the cylinder axis, relevant to elongation mechanism for bacteria shown in Fig. \ref{fig2}, the interactions fall off \textit{exponentially} at large distance  \cite{bib19,bib20}, due to a remarkable screening effect associated with vector charges, related to the physics of grain boundaries  \cite{bib5}.

\chapter {Third Sound:  van der Waals waves in thin helium films}

\begin{figure}
\includegraphics[width=0.6\textwidth, angle=-90]{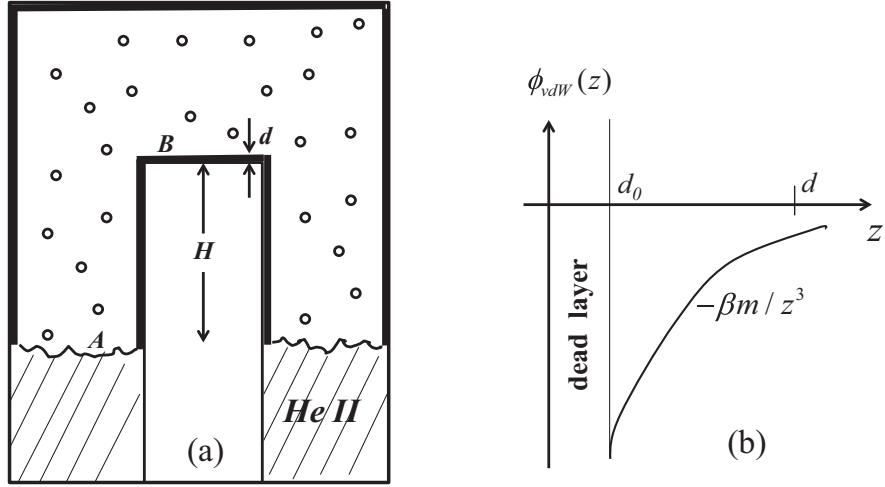}

\caption{(a) A pool of bulk superfluid helium ($He{\rm \; }II$) in closed chamber with cylindrical symmetry coats all walls with a superfluid helium film, including the cylindrical pillar in the center. The pillar projects a height $H$ above the surface of the bulk liquid, and the film has thickness $d$.  (b)   Van der Waals interaction energy $\phi_{vdW}(z)$ for a helium atom at height $z$ above a substrate.   This diverging energy leads to very high pressures for small $z$, causing a “dead layer” of inert non-superfluid helium atoms to form at distances less than $d_0$ from the substrate.}
\label{fig3}
\end{figure}

Below the lambda temperature $T_{\lambda }=2.17{}^\circ $Kelvin, liquids composed of $He^{4} $, the boson isotope of helium, flow without resistance between the narrowest parallel plates, through the smallest capillary tubes and through tightly packed powders like jewelers rouge.  Unless extreme care is taken, these fluids can easily defy gravity and climb out of their confining vessels  \cite{bib21}.   Remarkably, this superfluid phase (also denoted $He{\rm \; }II$), fails to crystallize and remains a liquid down to absolute zero at atmospheric pressure. At temperatures below $T_{\lambda } $, the dynamical properties of the superfluid phase can be understood in terms of a ``two-fluid model'', consisting of a normal fluid density $\rho _{n} (T)$, with a conventional shear viscosity, and a superfluid density $\rho _{s} (T)$, for which this viscosity is entirely absent  \cite{bib31}.   $He{\rm \; }II$ exploits its remarkable ability to flow without resistance by wetting nearly all substrates, with the exception of materials with very weak van der Waals interactions such as Cesium  \cite{bib32,bib33}.  Thus, as illustrated schematically in Fig. \ref{fig3}, a pool of bulk superfluid liquid helium in a closed chamber (in equilibrium with its non-superfluid vapor phase) will coat all walls with a superfluid film, including the cylindrical pedestal in the center.

If we focus for the moment on the thin film on the circular top of the pedestal at height $H$ above the bulk fluid, the normal fraction will be completely pinned by its viscosity, due to the close proximity of the substrate.  The normal fluid density will, in any case, typically be much less than the superfluid density in the low temperature limit considered here.   The superfluid velocity $\vec{v}_{s} (\vec{r},t)$ within this film will then obey the inviscid Navier-Stokes (or Euler) equation,
\begin{equation} \label{GrindEQ__2_}
\frac{\partial \vec{v}_{s} }{\partial t} +(\vec{v}_{s} \cdot \vec{\nabla })\vec{v}_{s} =-\frac{1}{\rho _{s} } \vec{\nabla }p+\vec{f}_{ext} ,
\end{equation}
where $p(\vec{r},t)$ is the fluid pressure and $\vec{f}_{ext} (\vec{r})=-\vec{\nabla }\Phi (\vec{r})$ is a constant external body force due to a potential energy function $\Phi (\vec{r})$ acting on every parcel of fluid.  Incompressibility implies that $\vec{\nabla }\cdot \vec{v}_{s} =0$.  What is the thickness $d$ of this film?    In the limit $H \gg d$ indicated in Fig. \ref{fig3}, the potential energy $\Phi _{0} $ exactly at the top of the film at the point \textbf{\textit{B}}, relative to the point  \textbf{\textit{A}}, arises from a combination of the gravitational potential and the energy that binds helium atoms to the substrate,
\begin{equation} \label{GrindEQ__3_}
\Phi _{0} (H,d)=mgH-\frac{\beta m}{d^{3} }  ,
\end{equation}
where $m$ is the mass of a helium atom. The second term arises from the $1/r^{6} $ van der Waals attraction of a helium atom at the film-vapor interface to the substrate, integrated over the 3-dimensional half-space occupied by the pillar.   The constant $\beta $ depends on the polarizability  of the material that forms the pillar  \cite{bib21}.  Note that the zero of the gravitational energy is set at the height of the bulk liquid and the zero of the interaction with the substrate is for a film thickness $d=\infty $.   In equilibrium, the value of $\Phi _{0} $ at a point \textbf{\textit{B}} on the surface of the bulk fluid well away from the pillar must be exactly the same as at point \textbf{\textit{A}}, which leads to a prediction for the thickness of the film,
\begin{equation} \label{GrindEQ__4_}
d(H)=(\beta /gH)^{1/3} \approx 4\times 10^{-6} (1{\rm \; }cm/H)^{1/3} .
\end{equation}
where the numerical estimate follows from the value of $\beta $ for a glass substrate  \cite{bib27}.   The prefactors for other substrates are similar.  For a macroscopic pillar height $H\sim 1{\rm \; }cm$, the film will be only \textit{d} $\sim$ 200 helium layers thick!

As indicated in Fig. \ref{fig3}b, $\phi _{vdW} (z)$, the van der Waals contribution to the potential energy of a liquid parcel at height $z$ above the substrate has a short distance cutoff at a distance $d_{0} $.  For $d<d_{0} $, the substrate forces are so large that the helium atoms solidify into a ``dead layer'', in which superfluidity is absent.   A superfluid film thus only exists for a range of heights $z$ with $d_{0} <z<d$.   Upon inserting gravitational and substrate interaction energies into Eq. \eqref{GrindEQ__2_}, we find that the velocity field in the film on top of the pillar at height $z$ obeys
\begin{equation} \label{GrindEQ__5_}
\frac{\partial \vec{v}_{s} }{\partial t} +(\vec{v}_{s} \cdot \vec{\nabla })\vec{v}_{s} =-\frac{1}{\rho _{s} } \vec{\nabla }\left[p+\rho _{s} gz-\beta \rho _{s} /z^{3} \right]
\end{equation}
Consider the relative magnitudes of the two downward body forces, $\vec{f}_{g} =-g\hat{z}$ and $\vec{f}_{vdW} =-(3\beta /z^{4} )\hat{z}$ on the right hand side of Eq. \eqref{GrindEQ__5_}.   Upon taking $z\sim d(H)=(\beta /gH)^{1/3} $ , we have$f_{vdW} /f_{g} \approx 3\beta /d^{4} g=3H/d$.    Thus, although the gravitational energy helps set the height of the film, it is utterly negligible compared to the van der Waals forces when $H\gg d$.

\begin{figure}
\includegraphics[width=\textwidth]{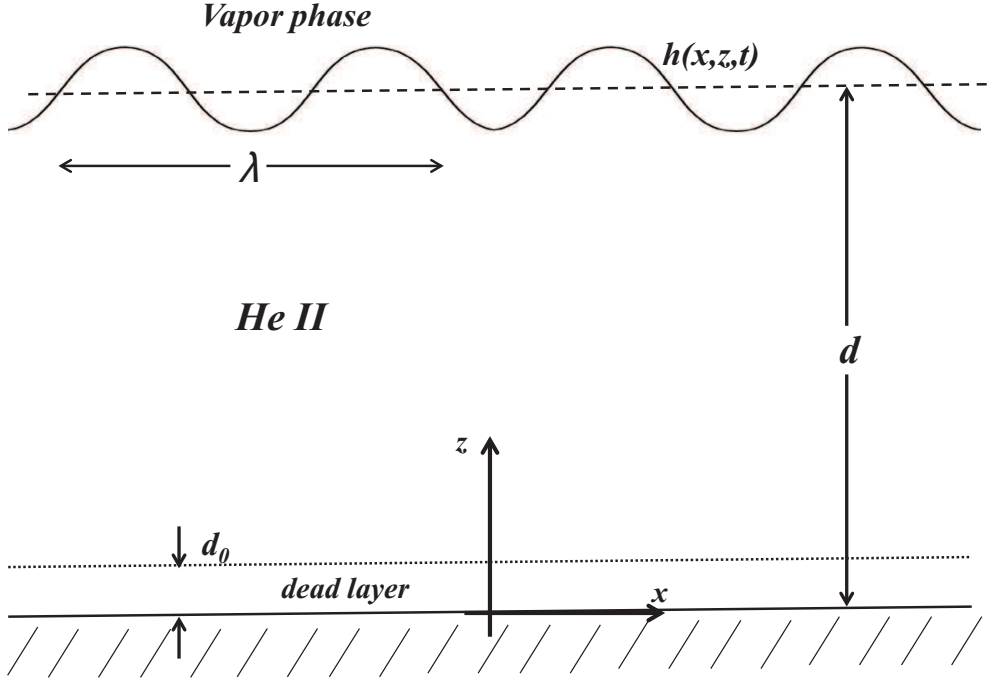}

\caption{ A van der Waals wave at the liquid-vapor interface of a superfluid helium film with equilibrium thickness $d$.   This excitation resembles a gravity wave, except that the dominant restoring force is the van der Waals interaction with the substrate.
    }
\label{fig4}
\end{figure}

 Consider the dynamics of the ripple-like third sound excitation at the liquid-vapor interface, as indicated schematically in Fig. \ref{fig4}.   This wave involves both a distortion of the interface, and a nonzero velocity field below the surface.   In Fig. \ref{fig4}, both the interface height $h(x,z,t)$ and the velocity field are independent of $y$.   As mentioned above, third sound is similar to a gravity wave in a classical fluid  \cite{bib34}, except that the restoring force of gravity replaced by interactions with the substrate.   Like gravity waves, the motion is \textit{potential flow,} \emph{i.e.}, $v_{s} (x,z,t)=\vec{\nabla }\phi (x,z,t)$.  However, unlike classical fluids, for important quantum mechanical reasons, the potential function is related to the phase $\theta (x,z,t)$ of the wave function that describes the macroscopic Bose-Einstein condensation that underlies the dissipationless superfluid  \cite{bib7,bib31}.    Indeed, at sufficiently low temperatures, we have
\begin{equation} \label{GrindEQ__6_}
\vec{v}_{s} (x,z,t)=\frac{\hbar }{m} \theta (x,z,t)\equiv \vec{\nabla }\phi (x,z,t),{\rm \; \; 0<}\theta <2\pi ,
\end{equation}
Where $m$ is the mass of a helium atom, and the restriction on the phase angle is natural because its relation to the condensate wave function is approximately $\psi _{0} (\vec{r})\approx \sqrt{\rho _{s} } e^{i\theta (\vec{r})} $  \cite{bib35}.   With the neglect of gravitational forces, Eq. \eqref{GrindEQ__5_} can thus be written
\begin{equation} \label{GrindEQ__7_}
\frac{\partial \vec{v}_{s} }{\partial t} =\frac{\partial }{\partial t} \vec{\nabla }\phi =-\frac{1}{\rho _{s} } \vec{\nabla }\left[p-\beta \rho _{s} /z^{3} \right].
\end{equation}
where we have neglected the nonlinear advective term $(\vec{v}_{s} \cdot \vec{\nabla })\vec{v}_{s} $.   Let us assume the excitation in Fig. \ref{fig4} has period $\tau $ and wavelength $\lambda $.  A criterion for the neglect of the nonlinear coupling follows if we first parametrize the interface height as
\begin{equation} \label{GrindEQ__8_}
h(x,t)=d+\zeta (x,t)\equiv d+\zeta _{0} e^{ikx-i\omega t}
\end{equation}
and then note that, $\partial v_{s} /\partial t\sim \zeta _{0} /\tau ^{2} $ whereas $(\vec{v}_{s} \cdot \vec{\nabla })v_{s} \sim \zeta _{0}^{2} /\tau ^{^{2} } \lambda $.   Our neglect of advection is thus justified for $\zeta _{0} \ll \lambda $, \emph{ i.e.} for excitations with wavelengths much longer than the interface amplitude $\zeta _{0} $.   From Eq. \eqref{GrindEQ__7_}, we have $\vec{\nabla }\left[\frac{\partial \phi }{\partial t} +\frac{p}{\rho _{s} } -\frac{\beta }{z^{3} } \right]=0$.   Upon neglecting an arbitrary function of time, which can absorbed into a redefinition of the velocity potential $\phi (x,z,t)$,  we conclude that
\begin{equation} \label{GrindEQ__9_}
p(x,z,t)=\frac{\beta \rho _{s} }{z^{3} } -\rho _{s} \frac{\partial \phi (x,z,t)}{\partial t} ,{\rm \; \; }d_{0} <z<d .
\end{equation}
Note that the first term on the right side leads to an extremely large pressure for small $z$, which drives the formation of the solid ``dead layer'' of helium atoms close to the substrate. Incompressiblity of the superfluid gives us an additional equation for the velocity potential within the film, namely
\begin{equation} \label{GrindEQ__10_}
\nabla ^{2} \phi (x,z,t)=0 .
\end{equation}

 Understanding third sound requires that we now specify the boundary conditions associated with Fig. \ref{fig4}.   An undulating interface gives rise to a restoring force due to van der Waals interactions with the substrate in thin films, analogous to the earth's gravitational field for gravity waves in the ocean.  Like oceanic gravity waves, it can be shown that the restoring force due to surface tension can be neglected at long wavelengths  \cite{bib34,bib36}. The velocity perpendicular to the wall of the inviscid fluid must vanish close to the dead layer, but some care is required with the upper boundary, which is itself a dynamical quantity, determined by the fluctuating interface position $d+\zeta (x,t)$.  Upon assuming film undulations so slow that the pressure in the vapor phase just above the film remains fixed at its equilibrium value $p_{0} $, we have from Eq. \eqref{GrindEQ__9_} that
\begin{equation} \label{GrindEQ__11_}
\frac{\beta \rho _{s} }{(d+\zeta )^{3} } -\rho _{s} \frac{\partial \phi (x,d+\zeta ,t)}{\partial t} =p_{0} ,
\end{equation}
which upon linearization in $\zeta $ leads to $\frac{-3\beta \rho _{s} }{d^{4} } -\rho _{s} \frac{\partial \phi (x,d,t)}{\partial t} =p_{0} -\frac{\beta \rho _{s} }{d^{3} } $.   The term on the right hand side can be eliminated by letting $\phi \to \phi -(t/\rho _{s} )(p_{0} -\beta \rho _{s} /d^{3} )$(this change of variables leaves $\vec{v}_{s} =\vec{\nabla }\phi $ unaffected), to obtain finally
\begin{equation} \label{GrindEQ__12_}
\frac{3\beta }{d^{4} } \zeta (x,t)+\frac{\partial \phi (x,d,t)}{\partial t} =0.
\end{equation}
A second equation associated with the free surface arises because the $z$-component of the velocity at the interfaces can be calculated as both $v_{z} =\left. \frac{\partial \phi (x,z,t)}{\partial z} \right|_{z=d+\zeta } \approx \frac{\partial \phi (x,d,t)}{\partial z} $ and $v_{z} =\frac{\partial \zeta (x,t)}{\partial t} $ , which implies that
\begin{equation} \label{GrindEQ__13_}
\frac{\partial \phi (x,d,t)}{\partial z} =\frac{\partial \zeta (x,t)}{\partial t}    .
\end{equation}
These two boundary conditions at the free surface can be combined into a single equation upon differentiating  Eq. \eqref{GrindEQ__12_} with respect to time.    We are left with a single set of equations for the velocity potential, namely
\begin{equation} \label{GrindEQ__14_}
\begin{array}{l} {\nabla ^{2} \phi (x,z,t)=0{\rm \; \; \; \; \; \; \; \; (bulk)}} \\ {\frac{3\beta }{d^{4} } \frac{\partial \phi (x,d,t)}{\partial z} +\frac{\partial ^{2} \phi (x,d,t)}{\partial t^{2} } =0{\rm \; \; \; (at\; the\; free\; surface)\; \; }} \\ {\frac{\partial \phi (x,d_{0} ,t)}{\partial z} =0{\rm \; \; \; (near\; the\;substrate)}} \end{array}
\end{equation}

 This unusual wave equation with a parabolic time-dependent boundary condition has the solution $\phi (x,z,t)=\phi _{0} \cosh (kz)e^{ikx-i\omega (k)t} $ in the limit $kd_{0} \ll 1$, with dispersion relation
\begin{equation} \label{GrindEQ__15_}
\omega (k)=\sqrt{\frac{3\beta }{d^{4} } k\tanh (kd)} \approx \sqrt{\frac{3\beta }{d^{3} } } k\equiv c_{3} k,{\rm \; \; }kd\ll 1.
\end{equation}

\noindent The basic features of the ``third sound'' excitations with speed $c_3$ described above have been well-verified in numerous experiments on helium films at low temperatures \cite{bib28,bib29}, provided one replaces Eq. \eqref{GrindEQ__15_} by
\begin{equation} \label{GrindEQ__16_}
\omega (k)\approx \sqrt{\frac{\rho _{s} }{\rho } \frac{3\beta }{d^{3} } } k\equiv c_{3} k,{\rm \; \; }kd\ll 1,
\end{equation}
where the factor $\rho _{s} /\rho $ that enters the third sound veloicity $c_{3}$ arises because only a fraction of the film is superfluid when the temperature $T>0$ \cite{bib27}. However, as the temperature gradually increases, a new source of dissipation damps out third sound waves, until they no longer propagate  \cite{bib29,bib30}.

\noindent

\chapter{ Superfluid vortices:  analogy with 2d electrostatics}

\noindent The new source of dissipation for third sound/van der Waals waves in helium films (which sets in well below the bulk superfluid transition temperature) is associated with a breakdown of the irrotational flow assumption embodied in Eq. \eqref{GrindEQ__6_}.    If we were analyzing gravity or capillary waves in a classical fluid, we could simply study the viscous dissipation associated with adding a continuous vorticity field to the potential flow discussed above.   Vorticity can arise in superfluid helium films as well, but it is \textit{quantized}.  Indeed, the connection of the superfluid velocity with the phase of a single-valued quantum-mechanical wave function (see Eq. \eqref{GrindEQ__6_}) leads immediately to
\begin{equation} \label{GrindEQ__17_}
\oint _{C}\vec{v}_{s} (\vec{r})\cdot d\vec{l}=\frac{\hbar }{m} \oint _{C}\vec{\nabla }  \theta (\vec{r})\cdot d\vec{l}=\frac{\hbar }{m} 2\pi s,{\rm \; }s=0,\pm 1,\pm 2,...
\end{equation}
If an array of vortices with topological ``charges'' $\{ s_{j} \} $ occupy a set of points $\{ \vec{r}_{j} \} $, this condition on the velocity field in an infinite plane leads to a nonzero 2d vorticity $\omega (\vec{r})$, namely
\begin{equation} \label{GrindEQ__18_}
\omega (\vec{r})=\hat{z}\cdot [\vec{\nabla }\times v_{s} (\vec{r})]=2\pi \frac{\hbar }{m} \sum _{j}s_{j} \delta ( \vec{r}-\vec{r}_{j} ).
\end{equation}

\noindent For a \textit{single} vortex with $s=1$ located at $\vec{r}_{0} =(x_{0} ,y_{0} )$ we have $\theta (x,y)=\tan ^{-1} \left(\frac{y-y_{0} }{x-x_{0} } \right)$, so that
\begin{equation} \label{GrindEQ__19_}
v_{x} (\vec{r})=-\frac{\hbar }{m} \frac{(y-y_{0} )}{|\vec{r}-\vec{r}_{0} |^{2} } ,{\rm \; }v_{y} =\frac{\hbar }{m} \frac{(x-x_{0} )}{|\vec{r}-\vec{r}_{0} |^{2} } ,
\end{equation}
a velocity field that is curl-free except at $\vec{r}=\vec{r}_{0} $.  Insertion of this velocity field into the superfluid kinetic energy displayed in the second column of Table I, leads to a logarithmically diverging energy, $E=\pi \rho _{s} (\hbar /m)^{2} \ln (L/a)$ ,where $L$ is the in-plane size of film, and $a$ is a microscopic cutoff, of order the spacing between helium atoms.    Henceforth, we restrict our attention to charge \textit{neutral} vortex pairs, which have a finite interaction energy.

A detailed consideration of the damping of third sound by point-like vortex pairs in helium films (there is an interesting analogy with Maxwell's equations  \cite{bib26}) would take us well beyond the scope of these lectures.   To motivate our investigation of vortices on cylinders, we discuss here only the energetics of vortex pairs in the presence of a \textit{uniform} supercurrent $\vec{v}_{s}^{0}$.   Although a close approximation to this background superflow can be created via a torsional oscillator experiment  \cite{bib38}, such a situation also arises as the long-wavelength limit of the third sound waves discussed above:   At wavelengths very much greater than the film thickness, the in-plane velocity field associated with the slowly-varying third sound wave looks approximately \textit{uniform} with magnitude $v_{s}^{0} $ to a tightly-bound vortex pair, similar to, say, electron-hole pairs in a semiconductor exposed to electromagnetic radiation at optical wavelengths much larger than the pair separation  \cite{bib1}.

Energy calculations for helium films are facilitated by an analogy with 2d electrostatics, as first pointed out and exploited in the seminal work of Kosterlitz and Thouless  \cite{bib29}. Kosterlitz and Thouless showed how defects such as dislocations and vortices could give rise to phase transitions in two-dimensional magnets, superfluids and crystals.   Their work is directly applicable to the vortex-unbinding transition that causes the vanishing of the superfluidity density with a universal jump discontinuity in helium films  \cite{bib40} at a two-dimensional critical temperature $T_{c} $, now called the Kosterlitz-Thouless transition.   We will use similar ideas to determine how vortex pairs polarize in the presence of uniform supercurrents when $0<T \ll T_{c} $.    First define a fictitious electric field $\vec{E}(\vec{r})=\hat{z}\times \vec{v}_{s} (x,y)$, which is simply a $90{}^\circ $ counterclockwise rotation of the velocity field.    Eq. \eqref{GrindEQ__18_} then becomes a two-dimensional version of Gauss' law,
\begin{equation} \label{GrindEQ__20_}
\vec{\nabla }\cdot \vec{E}(\vec{r})=2\pi \frac{\hbar }{m} \sum _{j}s_{j} \delta ( \vec{r}-\vec{r}_{j} ).
\end{equation}
At long wavelengths, the coarse-grained 2d in-plane velocity field $\vec{v}_{s} (x,y)$, averaged over the film thickness, greatly exceeds the out-of-plane component, so that we now have a 2d incompressibility constraint, $\vec{\nabla }\cdot \vec{v}_{s} =\partial _{x} v_{s}^{x} +\partial _{y} v_{s}^{y} =0$.   In terms of the fictitious electric field defined above, this constraint reads
\begin{equation} \label{GrindEQ__21_}
\hat{z}\cdot (\vec{\nabla }\times \vec{E})=0,
\end{equation}
which leads us to define an electrostatic potential $\Phi (x,y)$ via $\vec{E}(x,y)=-\vec{\nabla }\Phi (x,y)$ that satisfies
\begin{equation} \label{GrindEQ__22_}
\nabla ^{2} \Phi (x,y)=-2\pi \frac{\hbar }{m} \sum _{j}s_{j} \delta ( \vec{r}-\vec{r}_{j} ).
\end{equation}
In the continuum limit, $\Phi (\vec{r})$ is related to the stream function of 2d fluid mechanics.

\begin{figure}
\includegraphics[width=\textwidth]{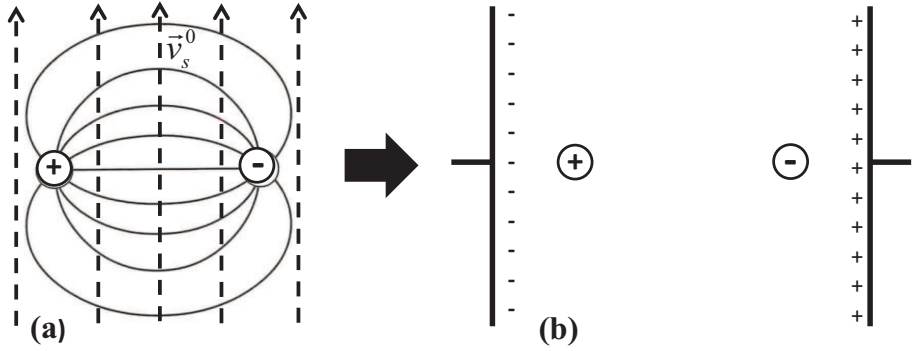}
\vspace{-3cm}
\caption{(a) Vortex dipole with lines of constant phase in a helium film in the presence of a uniform vertical supercurrent $\vec{v}_{s}^{0}$ (dashed lines).   (b)  The equivalent two-dimensional electrostatic problem consists of a pair of charges immersed a uniform horizontal electric field due to a charged capacitor, as the plate separation is taken to infinity.}
\label{fig5}
\end{figure}

As illustrated in Fig. \ref{fig5}, this analogy with electrostatics maps computing the kinetic energy of the velocity configuration associated with a vortex pair immersed in a uniform supercurrent, which can be written as an integral over the intensity of the ``electric field''
\begin{equation} \label{GrindEQ__23_}
U=\frac{1}{2} \rho _{s} \int d^{2} r|\vec{v}_{s} |^{2} = \frac{1}{2} \rho _{s} \frac{\hbar ^{2} }{m^{2} } \int d^{2} r\left[\vec{E}(\vec{r})\right] ^{2} =\frac{1}{2} \rho _{s} \frac{\hbar ^{2} }{m^{2} } \int d^{2} r\left[\vec{\nabla }\Phi (\vec{r})\right] ^{2}  ,
\end{equation}

\begin{figure}
\includegraphics[width=0.7\textwidth]{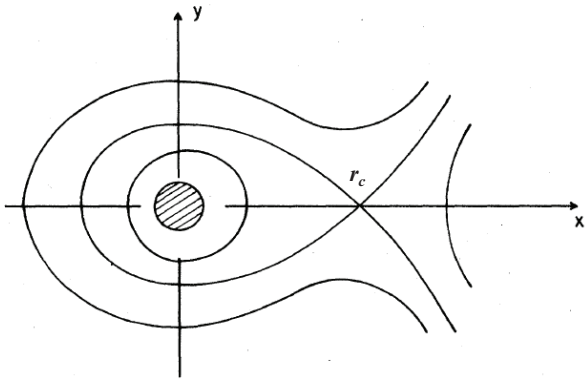}

\caption{(Adapted from Ref. [26])  Contours of constant potential between a pair of oppositely charged vortices at separation $r = (x,y)$. The potential decreases upon approaching origin, but has a short distance cutoff at radius $a$.   A uniform superfluid velocity $\vec{v}_{s}^{0}$ has been imposed in the $y$ direction.  There is a saddle point at $(r_c, 0)$, over which the pair separation can "escape" to large positive $x$.
}
\label{fig6}
\end{figure}

\noindent onto the 2d electrostatics problem of two charges in a uniform electric field proportional $\vec{z}\times \vec{v}_{s}^{0} $.    Let us define a ``quantum of vorticity flux'' $\omega _{0} =2\pi \hbar /m$.   For two vortices located at $\vec{r}_{1} $ and $\vec{r}_{2} $ with unit ``charges" $s_{1} =+1$ and $s_{2} =-1$, the electrostatic potential is
\begin{equation} \label{GrindEQ__24_}
\Phi (\vec{r})=\frac{-\omega _{0} }{2\pi } \ln (|\vec{r}-\vec{r}_{1} |/a)+\frac{\omega _{0} }{2\pi } \ln (|\vec{r}-\vec{r}_{2} |/a)-(\hat{z}\times v_{s}^{0} )\cdot \vec{r} .
\end{equation}
Upon inserting this potential function into Eq. \eqref{GrindEQ__23_}, we obtain, after integrations by parts and use of Eq. \eqref{GrindEQ__22_}, the energy function for the vortex pair,
\begin{equation} \label{GrindEQ__25_}
U(\vec{r}_{1} -\vec{r}_{2} )=const.+\rho _{s} \frac{\omega _{0}^{2} }{2\pi } \ln (|\vec{r}_{1} -\vec{r}_{2} |/a)-\rho _{s} \omega _{0} (\hat{z}\times \vec{v}_{s}^{0} )\cdot (\vec{r}_{1} -\vec{r}_{2} ),
\end{equation}
where the constant depends on the detailed physics near the vortex cores.    Although the logarithmic binding potential dominates when the vortex pair separation $r=|\vec{r}_{1} -\vec{r}_{2} | \ll r_{c} =\omega _{0} /(2\pi v_{s}^{0} )$, Fig. \ref{fig6} shows that the linear contribution to the interaction potential $U(\vec{r})$ leads to a saddle point along the $x$-axis at $\vec{r}=(r_{c} ,0)$.   Beyond this saddle point, the pair can separate indefinitely.   A detailed theory of this thermally activated ``escape over a barrier'' problem  \cite{bib26} (this theory now 1/3 of a century old!) reveals how long wavelength third sound waves are disrupted at low temperatures by this vortex unbinding process, with related results for the torsional oscillator experiments of Ref.  \cite{bib38}.    It is not our purpose here to recapitulate this ancient theory of vortex unbinding in planar helium films.  Rather, we conclude these lectures by discussing how Eq. \eqref{GrindEQ__25_} is altered, when we consider helium films on a cylinder.

\noindent

\chapter {Superfluid vortices on a cylinder}

\begin{figure}
\includegraphics[width=\textwidth]{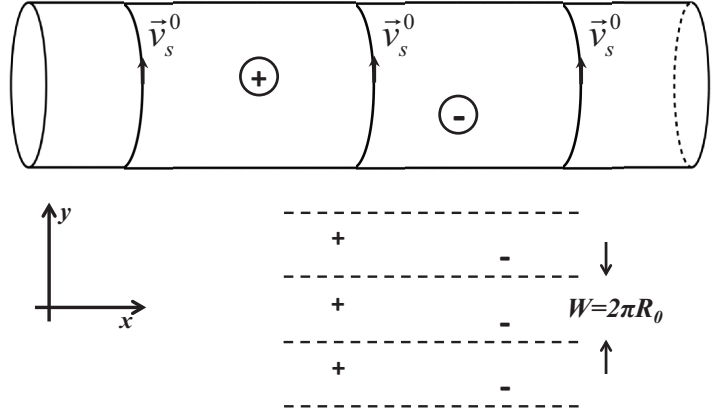}
\vspace{-3cm}
\caption{Top:  Vortex pair excitation in the presence of a constant background supercurrent $\vec{v}_{s}^{0}$ that wraps around the cylinder.   There is a force at right angles to this supercurrent (often called the Magnus force in fluid mechanics texts) that drives the vortices to separate towards the ends of the cylinder.    The response to more general supercurrents that spiral around the cylinder is discussed in the text.    Bottom right:   periodic replicas of the positive and negative vortices required by the cylindrical boundary conditions. }
\label{fig7}
\end{figure}

Inspired in part by the interacting dislocation defects on cylinders discussed in the introduction, we now ask how quantized vortex pairs interact in the presence of a uniform supercurrent in a cylindrical geometry.   What is the effect of the periodic boundary conditions, especially when the cylinder is long and narrow?   A uniform supercurrent on a cylinder could be generated experimentally by first coating a solid spindle with a helium film, and then spinning up this cylinder at constant angular velocity at temperatures \textit{above} superfluid phase transition, so that the fluid viscosity initially forces the film to follow the motion of the substrate.   If the rotating spindle is subsequently cooled well below the transition, and the rotation stopped, the superfluid fraction of the film will continue to rotate indefinitely  \cite{bib22}.   The most important decay mechanism for this supercurrent will be nucleation and separation of vortex pairs, as illustrated in Fig. \ref{fig7}.   Indeed, separation of a +/- vortex pair to infinity along the cylinder axis causes a $2\pi $ phase slip in the condensate order parameter one moves around the cylinder  \cite{bib22,bib26,bib44}, which reduces the background superfluid velocity, as discussed later in this section.

\begin{figure}
\includegraphics[width=0.85 \textwidth]{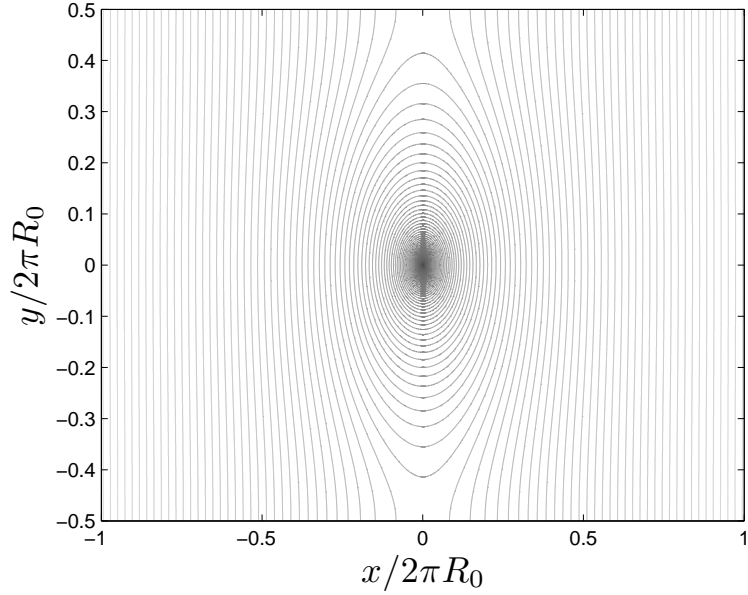}

\caption{Equipotential contours for a positive vortex with unit charge on a cylinder, interacting with a negative vortex at the origin. The interaction is logarithmic for $x \ll 2 \pi R_0$. However, for $x  \gg 2 \pi R_0$, the energy contours are parallel to the $y$ axis and reflect a linear, as opposed to a logarithmic, binding potential.}
\label{fig8}
\end{figure}

We first treat the interaction of a vortex pair in the absence of a supercurrent.   Suppose a positive vortex charge is glued to the origin, and its negative counterpart sits at position $\vec{r}=(x,y)$. We take the $y$-axis in the azimuthal direction, periodically repeated with period $W=2\pi R_{0}$ around the cylinder, and assume that the cylinder is infinite in the $x$-direction.  We need to solve Eq. \eqref{GrindEQ__22_}, subject now to the periodic boundary conditions imposed by the cylinder.     In the infinite plane, the energy would be $U(r)=U_{0} \ln (r/a)$, where  $U_{0} =\rho _{s} \omega _{0}^{2} /(2\pi )$, with a corresponding attractive force exerted on the + vortex given by $\vec{F}=-U_{0} \vec{r}/r^{2} $.    The generalization of this force to a cylinder requires summing an infinite number of negative image charges,
\begin{equation} \label{GrindEQ__26_}
F_{x} =-U_{0} \sum _{n=-\infty }^{+\infty }\frac{x}{x^{2} +(y+nW)^{2} } , {\rm \; \; \; }F_{y} =-U_{0} \sum _{n=-\infty }^{+\infty }\frac{y+nW}{x^{2} +(y+nW)^{2} }  .
\end{equation}
(The forces due to the positive image charges cancel, as they must so that the positive charge at the origin does not exert a force on itself.)  Similar sums over Matsubara frequencies are required to compute the Green's functions that arise quantum many body theory at finite temperatures  \cite{bib41}, and in many other fields.     They can be carried out exactly using the Sommerfeld-Watson formula, which reads  \cite{bib42}:
\begin{equation} \label{GrindEQ__27_}
\sum _{n=-\infty }^{\infty }f(n)= \frac{1}{2i} \oint _{C}f(z)\cot (\pi z)dz   ,
\end{equation}
where the clockwise contour $C$ for this complex contour integral encloses the entire real axis.  The proof of this formula (valid for functions $f(z)$ that fall off sufficiently rapidly for large $z$) proceeds by noting that the function $\cot (\pi z)$ inside the integrand has only simple poles of residue unity which lie on the x axis at integer values.   The sum is then evaluated by noting that since $f(z)\sim 1/|z|$ for large distances from the origin, we can deform the contour so that it captures only the poles of $f(z)$ in the complex plane  (note that the integral on the circle at infinity vanishes even though the decay is only $\sim 1/|z|$, due to the $\cot (\pi z)$ term.)   The results of this summation procedure for the forces are:
\begin{equation} \label{GrindEQ__28_}
\begin{array}{l} {F_{x} =\frac{1}{2} \frac{\pi U_{0} }{W} \left\{\coth [\pi (x-iy)/W]+\coth [\pi (x+iy)/W]\right\}}, \\ {F_{y} =-\frac{\pi U_{0} }{W} \frac{\sin (2\pi y/W)}{\cos (2\pi y/W)-\cosh (2\pi x/W)} }. \end{array}
\end{equation}
Note the periodicity in the $y$-coordinate.    Eq. \eqref{GrindEQ__28_} can be integrated to give the interaction energy for two vortices on a cylinder, namely
\begin{equation} \label{GrindEQ__29_}
U(x,y)=const.+\frac{1}{2} U_{0} \left[\ln \left\{\frac{W}{\pi a} \sinh [\pi (x+iy)/W]\right\}+c.c\right],
\end{equation}
a result originally obtained by Machta and Guyer  \cite{bib43}, who modeled helium films in porous media as a network of interconnected cylinders.   The contours of constant energy are displayed in Fig. \ref{fig8}.     It is easily checked that Eq. \eqref{GrindEQ__29_} reduces to the logarithmic interaction expected for the infinite plane when $|x|,|y| \ll W=2\pi R_{0} $.   More interesting, however, is the limit $x \gg W$, for which the forces in Eq. \eqref{GrindEQ__29_} read:
\begin{equation} \label{GrindEQ__30_}
\begin{array}{l} {F_{x} \approx \frac{\pi U_{0} }{W} {\rm sgn}(x)+\frac{2\pi U_{0} }{W} e^{-2\pi x/W} \cos (2\pi y/W),} \\ {F_{y} \approx \frac{2\pi U_{0} }{W} e^{-2\pi x/W} \sin (2\pi y/W).} \end{array}
\end{equation}
where $\rm{sgn}(x)$ is the sign of $x$.   Note the similarity to the electric field between two parallel capacitor plates, with exponential corrections due to the discreteness of the charges.   The cylindrical environment confines the electric field, so that it points predominantly along the cylinder axis.   Upon neglecting the exponential corrections and integrating, we see that the logarithmic binding potential for vortices is replaced by a linear binding for large vortex pair separations along the cylinder axis,
\begin{equation} \label{GrindEQ__31_}
U(x,y)\approx \frac{\pi U_{0} }{W} \left|x\right|,{\rm \; \; }\left|x\right| \gg W.
\end{equation}
However, the potential function that tends to \textit{separate} vortices in the presence of a circulating supercurrent like that in Fig. \ref{fig7} is \textit{also} linear in $x$!  (see Eq. \eqref{GrindEQ__25_}).    Hence, it is far from clear at this point that the escape-over-a-barrier scenario summarized in Fig. \ref{fig6} for an infinite planar film actually works for vortices on a cylinder, at least for small background supercurrents $\vec{v}_{s}^{0} $.   Which linear potential wins at large $x$, the repulsive one due to the supercurrent, or the attractive one due to electrostatics of point charges on a cylinder?    The remainder of this section provides a careful answer to this question, allowing for the more general case of a \textit{spiral} supercurrent $\vec{v}_{s}^{0} $ inclined at a finite angle to the $y$-axis.

The topological stability of supercurrents in the geometry of Fig. \ref{fig7} can be understood via the following argument:    If $\psi _{0} (x,y)$ is the single-valued condensate wave function describing to a superfluid helium film at rest with respect to a cylinder, adiabatically turning on a superfluid velocity $\vec{v}_{s} =(v_{s}^{x} ,v_{s}^{y} )$ (which might in general be inclined at a finite angle to the $y$-axis) leads to a new wave function
\begin{equation} \label{GrindEQ__32_}
\psi (x,y)=\exp [im\vec{v}_{s} \cdot \vec{r}/\hbar ]\psi _{0} (x,y),
\end{equation}
where $\vec{r}=(x,y)$and $m$ is the mass a helium atom.  For notational simplicity, we now drop the superscript $0$ on the background superfluid velocity, and set $\vec{v}_{s}^{0} \equiv \vec{v}_{s}$. The new wave function $\psi (x,y)$ must be single-valued as well;  hence, we have $\psi (x,y+W)=\psi (x,y)$.  The choice $\vec{v}_{s} \parallel \hat{x}$ represents a supercurrent along the cylinder axis which is \textit{not} quantized for an infinitely long cylinder.   However, the choice $\vec{v}_{s} \parallel \hat{y}$ describes an azimuthal supercurrent which must be quantized, according to $v_{s}^{y} =\frac{\hbar }{m} \frac{2\pi }{W} p,{\rm \; \; }p=0,\pm 1,\pm 2,...$.  Hence, a general supercurrent must take the form
\begin{equation} \label{GrindEQ__33_}
\vec{v}_{s} =v_{s}^{x} \hat{x}+\frac{\hbar }{m} \frac{2\pi }{W} p\hat{y},{\rm \; \; }p=0,\pm 1,\pm 2,....
\end{equation}
Thus the allowed values in the $(v_{s}^{x} ,v_{s}^{y} )$-plane are a set of horizontal lines with spacing $\frac{\hbar }{m} \frac{2\pi }{W} $.   As discussed in Refs. \cite{bib22} and  \cite{bib26}, vortex pair unbinding events mediate transitions between various states of superflow.    A general supercurrent will spiral around the cylinder, leading to a set of complementary spiral trajectories for the optimal unbinding of  vortex pairs.

Is there, similar to Fig. \ref{fig6} in the infinite plane, \textit{always} a saddle point for the escape-over-a-barrier energy landscape representing a vortex pair superimposed on a nonzero, uniform spiral background supercurrent?  Or, as one might na\"ively guess,  does a set of very small but nonzero supercurrents exist such that that the repulsive potential due to the background ``electric field" cannot overcome the electrostatic linear potential for a vortex pair that tries to separate along the cylinder and the vortex pair remains forever bound?   To answer this question,we need to understand the cylindrical energy landscape that generalizes Eq. \eqref{GrindEQ__25_}, namely
\begin{equation} \label{GrindEQ__34_}
U(x,y)=\frac{1}{2} U_{0} \left[\ln \left\{\frac{W}{\pi a} \sinh [\pi (x+iy)/W]\right\}+c.c\right]-\rho _{s} \omega _{0} \hat{z}\cdot (\vec{v}_{s} \times \vec{r}),
\end{equation}
subject to the quantization condition \eqref{GrindEQ__33_} above.   We have suppressed a short distance constant contribution that can be incorporated into the vortex core energies.    It is straightforward to analyze two limiting cases:

\noindent \textbf{\underbar{Case 1}:}  $\vec{v}_{s} \parallel \hat{x}$ (no quantization of $v_{s}^{x} $ on the infinite cylinder). We have $\hat{z}\cdot (\vec{v}_{s} \times \vec{r})=v_{s} y$ and  assume a saddle point of the form $(0,y_{c} )$, found by minimizing $U(0,y)=U_{0} \ln [\frac{W}{\pi b} \sin (\pi y/W)]-\rho _{s} \omega _{0} v_{s} y$.   It is easy to see that $y_{c}$ is the solution of $\cot (\frac{\pi y_{c} }{W} )=\frac{m}{\pi \hbar } v_{s} W$, which exists for \textit{any} value of $v_{s} $.  The solution $y_{c} (v_{s} )$ has the following properties: (a)  as $v_{s} \to \infty ,y_{c} \to 0$, so that vortex unbinding is very easy; (b)  as $W\to \infty ,$ we recover the infinite plane result, $y_{c} =r_{c} =\hbar /mv_{s} =\omega _{0} /2\pi v_{s} $; and finally     (c) $\mathop{\lim }\limits_{v_{s} \to 0} y_{c} =W/2$.  This last result is to be expected, since vortex pairs can always unbind by going azimuthally around the cylinder, due to the periodic boundary conditions.    Thus, as we let $v_{s} \to 0$ along the \textit{x}-axis (continuously, because there is no quantization), we always have a saddle point, and the conventional picture of Refs.  \cite{bib22} and  \cite{bib26} for the gradually destruction of superfluidity by successive unbindings of vortex pairs will hold.

\noindent

\noindent \textbf{\underbar{Case 2}}:  $\vec{v}_{s} \parallel \hat{y}$.  We now have a quantized superfluid velocity: $v_{s}^{y} =\frac{\hbar }{m} \frac{2\pi }{W} p,{\rm \; \; }p=0,\pm 1,\pm 2,...$.   In this case $\hat{z}\cdot (\vec{v}_{s} \times \vec{r})=-v_{s} x$ and we assume a saddle point of the form $(x_{c} ,0)$.   Consider a $\vec{v}_{s}^{} $ directed along the \textit{negative} y-axis, which requires that we now minimize  $U(0,y)=U_{0} \ln [\frac{W}{\pi a} \sinh (\pi x/W)]-\rho _{s} \omega _{0} v_{s} x$.    Although there is no saddle point for $p=0$, there is in fact a saddle point for every nonzero value of $p$, given by the solution of $\coth (\frac{\pi x_{c} }{W} )=2p$, or explicitly by $x_{c} (p)=\frac{W}{2\pi } \ln \left(\frac{2p+1}{2p-1} \right)$.    Note in particular that $x_{c}^{\min } =x_{c} (p)|_{p=1} =\frac{W}{2\pi } \ln 3$.  For large $W$ we have, $\mathop{\lim }\limits_{p\to \infty } x_{c} (p)=\frac{W}{2\pi p}=\hbar /mv_{s} =\omega _{0} /2\pi v_{s} $, as in flat space .

The ``na\"ive'' conclusion that no saddle point exists for small but nonzero supercurrents in the $y$-direction on the cylinder arises from comparing two competing linear terms in the vortex pair potential,
\begin{equation} \label{GrindEQ__35_}
\mathop{\lim }\limits_{x\to \infty } U(x,0)\approx \left[\frac{U_{0} \pi }{W} -\rho _{s} \omega _{0} v_{s} \right]x
\end{equation}
which leads to a ``na\"ive'' critical current $v_{s}^{c} =\frac{U_{0} \pi }{\rho _{s} \omega _{0} W} =\frac{1}{2} \frac{\hbar }{m} \frac{2\pi }{W} $.    However, this is half the minimum allowed nonzero azimuthal supercurrent on a cylinder, and hence \textit{cannot} not be realized experimentally, due to the constraint of a single-valued wave function.

\noindent

\begin{figure}
\includegraphics[width=1\textwidth ]{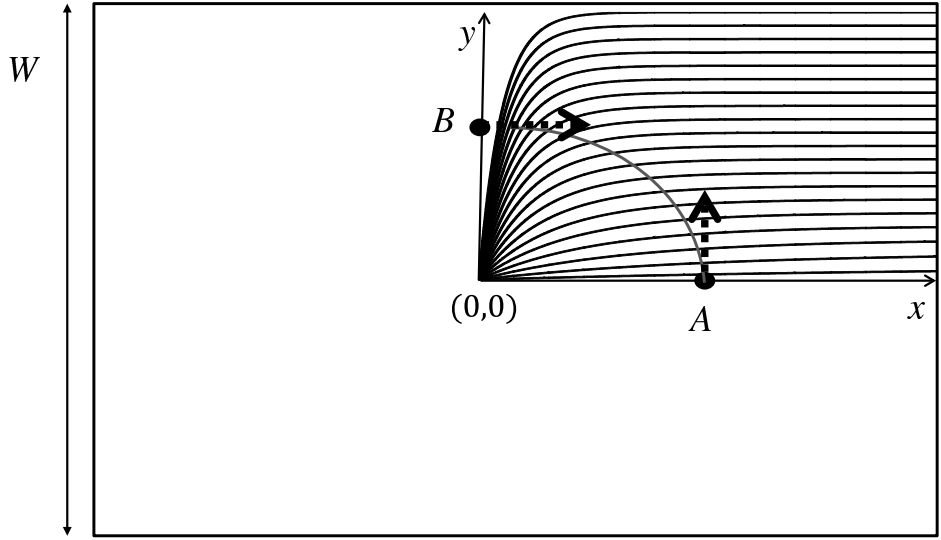}
\caption{A dense set of paths on the cylinder. The figure is an illustration used in the proof of the existence of a saddle point for any direction of the supercurrent. For each path the maximum of the potential energy is chosen, resulting in the line joining points $A$ and $B$. It can now be proved that there is a minimum on this curve, which is a saddle-point of the potential energy.  }
\label{fig9}
\end{figure}

The existence of a saddle point for cases 1 and 2 above strongly suggests that a saddle point exists for any choice of $\vec{v}_{s} $. This is indeed the case, as we shall now prove:   Since the potential energy of the vortices has an exact electrostatic analogy, it is clear that the potential energy is a harmonic function, and thus \textit{any} extremum has to be a saddle point. Let us assume without loss of generality that$v_{s}^{x} >0,v_{s}^{y} <0$. Using the arguments outlined above, we know that for $x\gg W$ and fixed \textit{y} the potential is approximately linear in \textit{x, }and the slope must be \textit{negative} since the argument relying on the quantization of $v_{s}^{y} $ still holds. Consider now a set of arbitrarily chosen paths (not to be confused with the energy contours or the unbinding trajectories!) each of which start at the origin and ends at $x\to \infty ,$ as shown in Fig. \ref{fig9}.  The potential energy near the origin is singular and attractive, so the slope as one moves from the origin towards positive $x$ is positive.  Thus, each of the paths has to have a maximum somewhere along it.  Moreover, from the properties of the potential we shall now show that each path must have a \textit{single} maximum along it: consider the \textit{x} component of the force that a vortex would encounter were it to be dragged along one of these paths. For any extremum this component must vanish. Along the path, the force component due to the supercurrent is constant, while the force due to the interaction with the vortex at the origin is monotonically decreasing (since it can be thought of as the superposition of the vortex and its image charges, all of which are monotonically decreasing). Therefore there cannot be more than a single maximum.

 For the lines \textit{x=0} and \textit{y=0,}  these maxima (points \textit{A} and \textit{B }in the figure) were previously discussed. Furthermore, we know that at point \textit{A} the force due to the vortex at the origin has no component in the \textit{y} direction (from symmetry), such that the gradient of the total potential energy is only contributed by the supercurrent and therefore points in the positive \textit{y} direction. Similarly, the gradient at point \textit{B }is in the positive \textit{x} direction. These are denoted by dashed arrows in Fig. \ref{fig9}. From continuity of the potential, and using the fact that there is a single maximum for each path, the set of maxima for all of the paths must lie on a single continuous path, denoted in the figure by the curve joining points \textit{A} and \textit{B}. Now let us consider the \textit{minimum} of the potential along this curve. It cannot lie on the boundary, since the gradients at the edges point inwards.  Therefore it must lie at some interior point. This point is thus a minimum in one direction, and a maximum in another -- and is therefore a saddle point.

\noindent

\chapter{Dislocations on cylinders are different}

\begin{figure}
\includegraphics[width=1\textwidth]{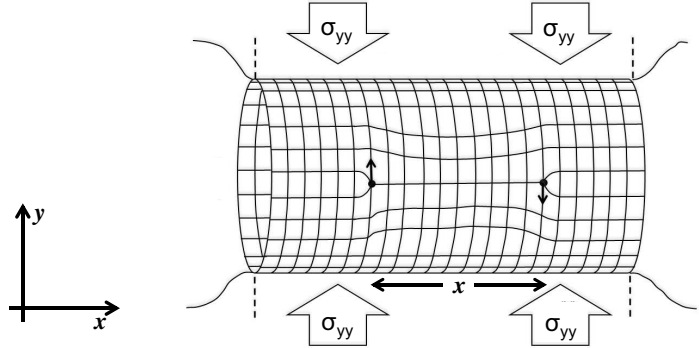}
\vspace{-4cm}
\caption{(adapted from Ref. [18])   Illustration of how dislocations could facilitate remodeling of a cylindrical bacterium growing through a narrow constriction bounded by the vertical dashed line segments. The cell wall assumes its natural radius $R_0$ of about $1\mu m$  outside the central region, which exerts a stress $\sigma_{yy}$ on the bacterium in the azimuthal direction.   If the dislocations are able to separate in the horizontal direction, the radius of the cell wall will shrink to accommodate the pore.}
\label{fig10}
\end{figure}

\noindent As discussed in Sec. I, dislocation defects in two dimensional elastic solids, whose close analogy with vortices in helium films is highlighted in table I, play an important role in in the elongation of the cell walls of cylindrical bacteria.  Fig. \ref{fig2} summarizes the model of bacterial elongation explored in Refs. \cite{bib18} and  \cite{bib19}.  Another example of 2d crystals with defects on a cylinder arises for interacting colloids on the surface of a liquid film coating a solid cylinder, where repulsive short-range forces give rise to the self-organized emergence of a two dimensional crystalline solid.   Defects in colloidal assemblies on the related curved surfaces of capillary bridges have been recently studied experimentally and theoretically  \cite{bib45}.   Here, the Gaussian curvature can be positive or negative; the zero Gaussian curvature of a cylinder is a special case  \cite{bib46}.    Fig. \ref{fig10} shows an alternative dislocation pair, with equal and opposite Burgers vectors $\pm b$ now parallel to the azimuthal direction.   This configuration might allow a bacterium to thread through a narrow circular pore smaller than its natural $\sim $1$\mu $m diameter, to which it relaxes on the left and right sides of the pore in the figure. The circular constriction produces a two-dimensional stress (force per unit length) $\sigma _{yy} $ that tends to drive the dislocations apart.    As discussed below, the dislocations are logarithmically bound for distances short compared to the circumference of the cylinder, similar to a $\pm $ vortex pair.     The indicated component $\sigma _{yy} $ of the stress tensor tends to squeeze the horizontal lattice rows together.   Like the vortex pair in the presence of a background supercurrent shown in Fig. \ref{fig7}, this force leads to an additional linear potential forcing the two dislocations apart.  Once these defects separate completely (for this geometry, an entire row of unit cells would have to be removed along the cylinder), the crystal would have reduced its circumference by one lattice constant.   Superficially, the dislocation configurations such as those shown in Figs. \ref{fig2} and \ref{fig10} seem quite similar to superfluid vortices on a cylinder.   Nevertheless, as highlighted below, there are important and surprising differences, due to the vector nature of the Burgers vector topological charges of the dislocations.   See Ref.  \cite{bib20} for a more thorough and extensive discussion of the physics of dislocations on cylinders summarized briefly below.
\noindent

In the infinite plane, the interaction energy of two edge dislocations with Burgers vectors $\vec{b}_{1} $ and $\vec{b}_{2} $, and with a relative separation $\vec{r}=\vec{r}_{1} -\vec{r}_{2} $, in the presence of a uniform background stress tensor $\sigma _{ij}$, is given by
\begin{equation} \label{GrindEQ__36_}
U(\vec{r})=const.+\frac{Y}{4\pi } \left[\vec{b}_{1} \cdot \vec{b}_{2} \ln (|\vec{r}|/a)-\frac{(\vec{b}_{1} \cdot \vec{r})(\vec{b}_{2} \cdot \vec{r})}{|\vec{r}|^{2} } \right]-b_{k} \sigma _{jk} \varepsilon _{ijz} r_{i}   ,
\end{equation}
where $a$ is a microscopic cutoff of order the lattice spacing and Y is an elastic constant, the 2d Young's modulus for the case of an isotropic crystal  \cite{bib7}.   The last term is the energy corresponding to the Peach-Koehler force  \cite{bib5,bib47} due to the stress field $\sigma _{jk}$, where $\varepsilon _{ijz} $ is the Levi-Civita tensor.   Comparison with Eq. \eqref{GrindEQ__25_} for the energy of a vortex pair with a background supercurrent shows that for crystalline solids, the Peach-Kohler force is the analog of the Magnus force on a vortex.   In a coordinate system where $\sigma _{ij} $ is diagonal ($\sigma _{xy} =0$), this force is $\vec{F}=-b\sigma _{xx} \hat{y}$ when $\vec{b}=b\hat{x}$ (as in Fig. \ref{fig2}) and $\vec{F}=b\sigma _{yy} \hat{x}$ when $\vec{b}=b\hat{y}$ (as in Fig. \ref{fig10}).

\noindent

\begin{figure}
\includegraphics[width=1\textwidth]{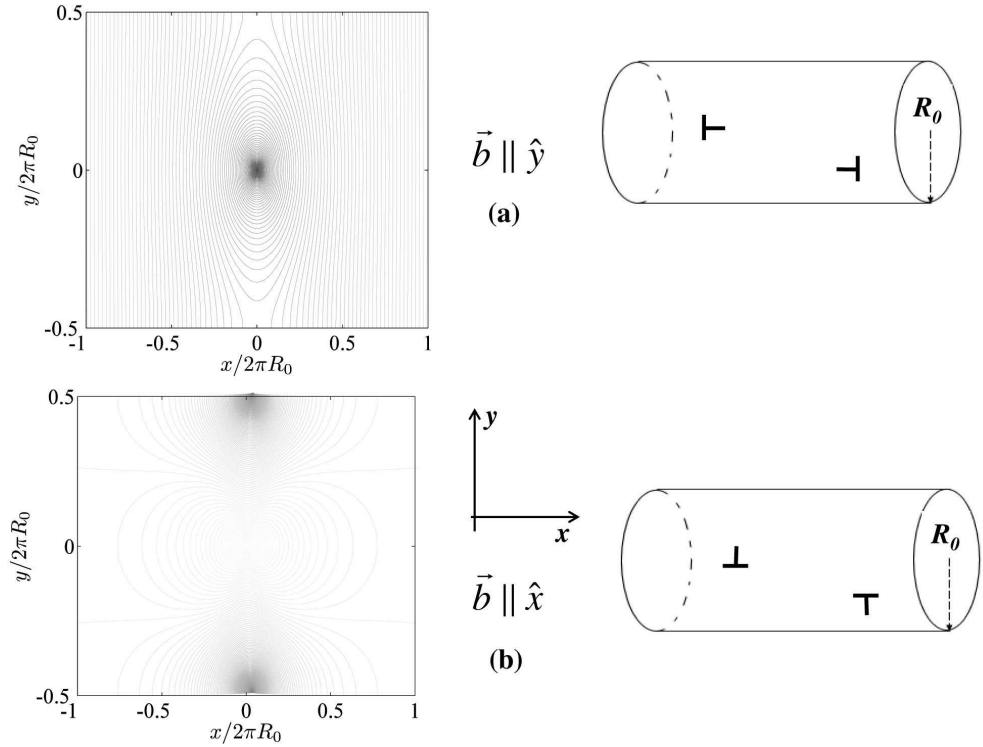}

\caption{ (a) Equipotential contours for an interacting dislocation pair, one at the origin and another at $(x, y)$, on a cylinder with azimuthal Burgers vectors similar to Fig. \ref{fig10}.    (b) Similar plot for a dislocation pair with Burgers vectors parallel to the cylinder axis, as in Fig. \ref{fig2}.   Unlike the linear potential at large horizontal separations in (a), the interactions are exponentially screened in this case.
}
\label{fig11}
\end{figure}

 On a cylinder, the potential energy of a dislocation pair takes the form
\begin{equation} \label{GrindEQ__37_}
U(\vec{r})=const.+E(\vec{b}_{1} ,\vec{b}_{2} ;x,y)-b_{k} \sigma _{jk} \varepsilon _{ijz} r_{i} ,
\end{equation}
where the linear potential due to a background stress is unchanged, and the interaction energy $E(\vec{b}_{1} ,\vec{b}_{2} ;x,y)$ is periodic in $y$ with period $W=2\pi R_{0} $ and can be calculated by summing over periodic image dislocations using the Sommerfeld-Watson method as sketched above for vortices  \cite{bib20}. Fig. \ref{fig11}a shows the energy contours for $E(b\hat{y},-b\hat{y};x,y)$, \emph{i.e.}, for dislocations oriented as in Fig. \ref{fig10}.   The potential is isotropic and logarithmic for short distances, and the dislocation pair experiences a linear potential at large separations along the x-axis, like vortices on a cylinder.   However, the energy contours for $E(b\hat{x},-b\hat{x};x,y)$, \emph{i.e.}, for dislocations oriented as in Fig. \ref{fig2}, differ dramatically, as shown in Fig. \ref{fig11}b.  The pair interaction, although isotropic and logarithmic for small separations, is now \textit{exponential screened} when $x \gg W$, \emph{i.e.} for separations much larger than the circumference  \cite{bib20}.  The source of this remarkable behavior is the dipole-like angular interaction term in Eq. \eqref{GrindEQ__36_} on the cylinder.    When $\vec{b}$ is parallel to $\hat{x}$, as is the case for the dislocations that mediate bacterial elongation in Fig. \ref{fig2}, the angular term causes the long range strain field from an isolated dislocation to fall off exponentially fast.  The reason for this strange behavior is that isolated dislocations on a cylinder with $\vec{b}||\hat{x}$ act like \textit{grain boundaries} in an infinite crystalline solid, due to the periodic boundary conditions.  Grain boundaries connect two orientationally mismatched crystals, and can be regarded as an infinite row of aligned dislocations with the same topological charge  \cite{bib5,bib7,bib48}.  The innocuous-looking angular term in Eq. \eqref{GrindEQ__36_} leads to a low energy cost for grain boundaries, contrary to what one might have guessed by a na\"ive application of the electrostatic analogy to a array of dislocations with identical charges.    The vector nature of these charges is crucial.

\noindent

Thus, despite the many deep and useful analogies between superfluid helium films and 2d crystalline solids, important differences emerge, due to the vector nature of the topological charges characterizing the dislocations.   See Ref.  \cite{bib20} for more details.

\acknowledgements
One of us (D. R. N.) is grateful to Lyd\'eric Bocquet, David Qu\'er\'e and Tom Witten for the opportunity to lecture once again in the beautiful setting of Les Houches.  We thank B. I. Halperin, W. T. M Irvine, F. Spaepen and V. Vitelli for useful discussions. This work was supported by the National Science Foundation via Grant DMR1005289 and through the Harvard Materials Research Science and Engineering Laboratory, through Grant DMR0820484. A.A. was supported by a Junior Fellowship of the Harvard Society of Fellows.


\thebibliography{0}
\bibitem {bib37} 37.  ``First sound'' is the usual pressure wave that arises in bulk superfluids when the normal and superfluid fractions oscillate in phase. ``Second sound'' allows wavelike propagation of heat and entropy in bulk superfluids, and arises when the normal and superfluid fractions oscillate in phase  \cite{bib31}.  ``Fourth sound'' is an excitation that arises in the porous materials such as bulk vycor glass  \cite{bib27}. It is carried almost entirely by the superfluid fraction (much like a third sound in thin superfluid films).
\endthebibliography
\end{document}